\documentclass[preprint,showpacs,preprintnumbers,amsmath,amssymb,eqsecnum,aps]{revtex4}

\usepackage{graphicx}
\usepackage{dcolumn}
\usepackage{bm}

\begin{document}


\title{Exact computation of the mean velocity, molecular diffusivity and dispersivity of a particle moving on a
periodic lattice}

\author{Kevin D. Dorfman}
 \email{Kevin.Dorfman@curie.fr}
 \affiliation{Institut Curie, Physico-Chimie/UMR 168, 26 Rue d'Ulm, 75248 Paris Cedex
5, France}

\date{\today}

\begin{abstract}
A straightforward analytical scheme is proposed for computing
the long-time, asymptotic mean velocity and dispersivity (effective diffusivity) 
of a particle undergoing a discrete biased random walk on a periodic lattice amongst
an array of immobile, impenetrable obstacles.  
The results of this Taylor-Aris dispersion-based theory are exact,
at least in an asymptotic sense, and furnish an
analytical alternative to conventional numerical lattice
Monte Carlo simulation techniques. 
Results obtained for an obstacle-free lattice are employed to establish
generic relationships between the transition probabilities, lattice size and 
jump time.
As an example, the dispersivity is computed for a solute moving through an isotropic 
array of obstacles
under the influence of a finite external field.
The calculation scheme is also
shown to agree with existing zero-field results, the
latter obtained elsewhere
either by first-passage time analysis or use of the Nernst-Einstein equation in the zero-field limit.
The generality of this scheme permits 
the study of more complex lattice structures, in particular trapping geometries.
\end{abstract}

\maketitle

\section{Introduction}\label{Intro}
Lattice Monte Carlo simulation constitutes a powerful method for computing the average transport rates
of a particle moving through a fixed network of obstacles by modeling the continuous transport process as a discrete 
random walk, biased or unbiased, 
across a lattice.  At the heart of this technique lies a master equation
\cite{vanKampen:81} which quantifies the probability of jumping from one point on the lattice to an adjacent
point in a given period of time.  
Owing to the stochastic (probabilistic) nature of the master equation, computer simulations
based upon lattice Monte Carlo techniques require numerous realizations of the system in order to
yield statistically significant results.  Moreover, since average transport rates are valid only for long
times, each simulation must be sufficiently long in order to satisfy this criteria.

Gel electrophoresis is one particularly relevant application of lattice Monte Carlo techniques,
where the fibers of 
the gel are modeled as obstacles in the network which ``reject'' solute jumps onto them as
a result of hard-core repulsion forces.  
In a recent series of publications \cite{Slater1,Slater2,
Slater3,Slater4,Slater5,Slater6,Slater7,Slater8,Slater9}, Slater and coworkers invoked lattice Monte
Carlo-type schemes to compute the effective solute mobility, $\bar{\mathbf{M}}^{\ast}$, 
for transport in spatially periodic model gels
in the ``Ogston-sieving'' regime of gel electrophoresis.
The latter works are unique in that they do not rely upon computer simulations at all. 
Rather, analytical solutions are 
obtained for the solute mobility, based upon the fact that, for periodic lattices, 
unit-cell probabilities can be computed analytically in
the long-time (steady-state) limit.
As a consequence, Slater and coworkers have been able to consider
a vast array of periodic model gels, both 2-D and 3-D, thereby 
critically assessing the validity of existing gel electrophoresis theories 
and investigating the effects of gel structure on the resultant average solute mobility.
Results obtained from these types of calculations are presumably exact, 
at least in an asymptotic sense, for perfectly
periodic structures, such as those which may be realized 
in microfluidic environments \cite{Chou:99,Cabodi:02,Doyle:02}.
Results for pseudo-random structures, which may be more representative of real gels, 
can be obtained by repeating the periodic 
calculation scheme numerous times in conjunction with random samplings of possible
obstacle configurations (using the spatial periodicity to 
mimic macroscopic-sized gels), and then
averaging the ensemble of results \cite{Slater3,Koplik:82}.  
The efficiency of analytical methods renders such calculations very feasible.

Although the aforementioned series of papers \cite{Slater1,Slater2,
Slater3,Slater4,Slater5,Slater6,Slater7,Slater8,Slater9} has made significant progress in addressing the
issue of the average solute mobility (or, equivalently, the mean solute velocity, $\bar{\mathbf{U}}^{\ast}$), 
much less progress has been made towards calculating 
solute {\em dispersion} arising as a result of the stochastic nature of the transport on the lattice.  
In the context of separation sciences \cite{Giddings:91}, 
it is essential to quantify both the dispersion and the average mobility,
since the relative mobility of the particles being separated determines 
the degree of the separation, while the dispersion
governs the separation sharpness.  Moreover, while
the local solute mobility dyadic, $\mathbf{M}$, and diffusivity dyadic, $\mathbf{D}$,
are related at each point in space by the Stokes-Einstein
equation \footnote{
In free solution, these dyadics are isotropic and the solute mobility reduces to its Stokes mobility.
},
\begin{equation}
\mathbf{D} = \mathbf{M} kT, \label{Stokes}
\end{equation}
with $kT$ the Boltzmann factor, the comparable Nernst-Einstein 
relationship between 
the dispersivity dyadic, $\bar{\mathbf{D}}^{\ast}$, and the average mobility dyadic,
$\bar{\mathbf{M}}^{\ast}$,
\begin{equation}
\bar{\mathbf{D}}^{\ast} = \bar{\mathbf{M}}^{\ast} kT, \label{Nernst}
\end{equation}
is only valid in the limit of vanishingly small applied fields.  

Mercier {\em et al.} \cite{Mercier:99} exploited this property of eq.  (\ref{Nernst})
to successively compute the components of $\bar{\mathbf{D}}^{\ast}$ 
 in the zero-field
limit \footnote{The zero-field limit dispersion coefficient
is often referred to as the diffusion coefficient for the
system, which should not be confused with the molecular diffusivity of the solute.
For clarity, we will only use the term diffusion in the context of local motion and the
term dispersion (or effective diffusion) in the context of averaged global motion.
}.
Explicitly, an analytical scheme
\cite{Slater1} was invoked to compute the different mean velocity vectors, say, $\bar{\mathbf{U}}^{\ast}_x$ 
and $\bar{\mathbf{U}}^{\ast}_y$,  
corresponding to infinitesimally small forces being applied in the subscripted
spatial directions.  The average mobility was then obtained by using the linear relationship,
\begin{equation}
\bar{\mathbf{U}}^{\ast} = \bar{\mathbf{M}}^{\ast} \cdot \mathbf{F}, \label{Mstar}
\end{equation}
in conjunction with each of the mean velocities.  The resulting average mobility tensor proves to be 
independent of field strength, whereupon eq.  (\ref{Nernst}) immediately furnishes the dispersivity dyadic,
$\bar{\mathbf{D}}^{\ast}$.
The resulting dispersivity was shown to agree exactly with that 
obtained by first-passage time analysis \cite{Mercier:99}.
Given the relative ease of the mobility calculation compared to its first-passage time counterpart,
Mercier {\em et al.} advocate the use of the former.

While calculating the dispersion (effective diffusion) coefficient for an unbiased random walk 
on the lattice is clearly a non-trivial endeavor,
this is the limiting case of the much broader problem of dispersion occurring during
{\em biased} random walks, the bias being induced by the presence of finite electric fields (or some other 
external force) \footnote{
By an unbiased walk, we mean that all jumps are equally likely at 
a given lattice site, but where some of the jumps may be rejected by the presence
of the obstacles.  In contrast, for a biased random walk, the likelihood of making
a jump in a particular direction is biased by the presence of the imposed force, resulting
in a preferred direction of motion.  Jumps onto sites occupied by the obstacles are
still rejected in the biased random walk.}.  
Keller {\em et al.} \cite{Keller:02} recently proposed a method for
computing the dispersivity dyadic from biased lattice walks in the low-field limit, 
based upon what is essentially a volume-averaging scheme.  In \S\ref{Mercier}, 
we will demonstrate that, in contrast to the present theory, the latter volume-averaged
scheme does not furnish the correct effective diffusivity for the anistropic array discussed
above \cite{Mercier:99}. While this does not address the underlying validity (or lack thereof)
of the volume-averaged approach, it does bring into question its applicability to the present class of problems.
Consequently, no simple method exists in the literature
for computing solute dispersion from a generic lattice model,
and it has been noted \cite{Slater9} that such a calculation scheme would be extremely useful for analyzing 
trapping-based separations.  

The present contribution addresses this need by developing
 an asymptotically exact, analytical scheme for computing 
the mean velocity vector, $\mathbf{\bar{U}}^{\ast}$, and dispersivity dyadic, 
$\mathbf{\bar{D}}^{\ast}$, from lattice Monte Carlo models,
by way of discrete generalized Taylor-Aris dispersion theory for spatially periodic
networks \cite{Adler:84b,Dorfman:02}.  The latter theory was initially developed as a means to 
compute $\mathbf{\bar{U}}^{\ast}$ and $\mathbf{\bar{D}}^{\ast}$
from lumped-parameter models of transport in model
porous media \cite{Adler:84b} and microfluidic networks \cite{Dorfman:02}.  We will show that 
the lattice random walk has much in common with the 
existing lumped-parameter transport problem, whereupon the moment scheme invoked previously \cite{Adler:84b,Dorfman:02}
may be readily employed here to homogenize the master equation governing transport on the periodic lattice.
Consideration of the obstacle-free case furnishes generic relationships between the jump time, transition 
probabilities and the lattice spacing.  
These generic relationships between the parameters are used in an illustrative example
of a solute moving on a simple
periodic lattice in the presence of a finite external 
field.
The theory simplifies substantially
in the absence of a field, leading to a compact scheme for computing the 
effective diffusion coefficient.
The simplified theory is invoked to compute the zero-field
dispersion coefficients 
for the asymmetric lattice problem originally considered
by Mercier {\em et al.} \cite{Mercier:99}, demonstrating that 
the present theory successfully 
reproduces their results.  

\section{Biased Random Walk on a Periodic Lattice}
\subsection{Geometry}
\begin{figure}[t]
\includegraphics[clip,width=2.5in]{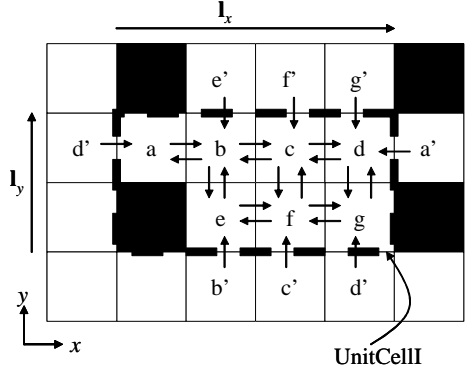}
\caption{
Basic graph, $\Gamma_b$, of a periodic lattice with an asymmetric array of obstacles.
The nodes of the graph are indicated by the lettered white boxes ($a$-$g$), 
and the edges of the graph are indicated by the arrows.  The periodic unit cell is indicated by the heavy dashed line, and those nodes with a prime affix
correspond to homologous vertices lying outside the unit cell.
Base lattice vectors $\mathbf{l}_x$ and $\mathbf{l}_y$ are indicated.  
Only those edges entering the unit cell are included in the basic graph.  Diffusion in this particular array
was previously considered in Ref. \onlinecite{Mercier:99}.
\label{Fig1} 
}
\end{figure}

Consider a periodic lattice of the type depicted in Fig. \ref{Fig1}.  The available sites of the lattice
are indicated by white boxes and the unavailable sites, corresponding to the obstacles in the array, are
indicated by black boxes.  Transport between lattice sites occurs by discrete ``jumps'' 
between them.
The spatial periodicity of the lattice is reflected by the presence of a repetitive
unit cell, which reproduces the entire unbounded medium upon translation by its 
base lattice vectors $(\mathbf{l}_1,\mathbf{l}_2,\mathbf{l}_3)$.
Each cell is characterized by its respective discrete location vector, $\mathbf{I} = (I_1,I_2,I_3)$, with the origin 
arbitrarily chosen to be located at the cell $\mathbf{I_{0}}=\mathbf{0}$.  The position of the locator point, say,
the centroid, of cell $\mathbf{I}$ is given by the discrete vector $\mathbf{R_I} = I_1\mathbf{l_1} + 
I_2\mathbf{l_2} + I_3\mathbf{l_3}$.

The unbounded, periodic lattice may be represented by 
a series of three graphs (although only the unit-cell based 
local graph, $\Gamma_l$, will be necessary
for computing $\bar{\mathbf{U}}^{\ast}$ and $\bar{\mathbf{D}}^{\ast}$),
where the nodes of the graph correspond to the lattice sites and the edges of the 
graph correspond to the possible jumps between sites.  
The specifics of the graphical construction are discussed at length in Ref.  \onlinecite{Dorfman:02},
whereupon the following exposition will be brief.  
Each lattice site (node) is assigned a discrete
location $(\mathbf{I},i)$, where $i$ represents the intracellular position (i.e. $a$-$g$ in Fig. \ref{Fig1}).
As a consequence of the spatial periodicity of the network, 
it only proves necessary to consider edges of the type indicated in
Fig. \ref{Fig1}. Let $\Omega^{+}(i)$ be the subset of 
the edges $j$ entering node
$(\mathbf{I},i)$ from node $(\mathbf{I'},i')$, and $\Omega^{-}(i)$ be those edges exiting node $(\mathbf{I},i)$
and entering node $(\mathbf{I'},i')$ \footnote{
For those edges contained entirely in unit cell $\mathbf{I}$, $\mathbf{I}=\mathbf{I'}$.
}.  The global graph, $\Gamma_g$, contains all of the infinitely many nodes 
and all of the edges connecting them, equipollent with the 
unbounded lattice.  As an intermediate step, we construct the basic graph, $\Gamma_b$, 
depicted in Fig. \ref{Fig1}, which includes: 
(i) all of the nodes in the unit cell; (ii) all those edges contained wholly within the unit cell;
(iii) those
edges {\em entering} the unit cell; and (iv) their associated homologous vertices, 
the latter indicated with the prime affix in 
Fig. \ref{Fig1}.  

A macroscopic jump vector, 
\begin{equation}
\mathbf{R}(j) = \mathbf{R_I} - \mathbf{R_{I'}}, \label{Rj}
\end{equation}
is assigned to each of the edges, this vector being identically zero for those edges which do not
cross the boundary of the unit cell.  
 This permits one to distinguish between edges connecting, say, $e' \rightarrow b$ and
$e \rightarrow b$ in the basic graph, since their respective macroscopic jump vectors are
$\mathbf{R}=-\mathbf{l}_y$ and $\mathbf{R}=\mathbf{0}$.  Use of the macroscopic jump
vectors prevents some (potentially significant) geometrical simplifications achieved in the 
lattice models proposed by Slater and coworkers 
\cite{Slater1,Slater2}, since ``identical'' sites in their model now typically differ
with respect to the macroscopic jump vectors in $\Omega^{+}(i)$.
Since the velocity may be computed without
use of the these vectors [cf. eq. (\ref{SlaterU})], their scheme \cite{Slater2} for computing
$\bar{\mathbf{U}}^{\ast}$ is more efficient
for certain geometries.
However, the primary contribution here is a 
a scheme for 
computing $\bar{\mathbf{D}}^{\ast}$ [cf. eq. (\ref{Dstar})], which  
requires using the macroscopic jump vectors to compute the moments of the probability
[cf. eq. (\ref{localdef})].

\begin{figure}[t]
\includegraphics[clip,width=2.5in]{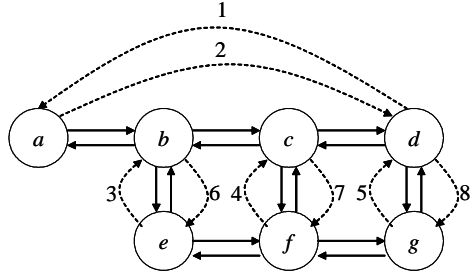}
\caption{
The local graph, $\Gamma_l$, constructed from the basic graph of Fig. \ref{Fig1}.  Those edges with non-zero
macroscopic jump vectors $\mathbf{R}(j)$ are indicated by the dashed lines.  Edge numbers
refer to the calculation of \S \ref{Mercier}.  \label{Fig2}
}
\end{figure}

The local graph, $\Gamma_l$, depicted in Fig. \ref{Fig2}, is constructed by combining the homologous vertices
on $\Gamma_b$
and contracting those edges between them.  The dashed lines in Fig. \ref{Fig2} correspond to edges
which cross the boundary of the unit cell in the basic graph (and thus have non-zero macroscopic
jump vectors).  
Reference to the discrete position vector $\mathbf{I}$ is no longer 
necessary in the local graph, since the periodic structure of the medium has been embedded in the macroscopic
jump vectors. 

\subsection{Master Equation}
Consider the probability $P(\mathbf{I},i,N \,|\, \mathbf{I}_0,i_0,0)$ that
the particle will be located at site $(\mathbf{I},i)$ on $\Gamma_g$ at step $N$, given its initial location at site 
$(\mathbf{I}_0,i_0)$.  In the usual manner \cite{Dorfman:02}, the final results only depend upon the 
displacement, $\mathbf{I}-\mathbf{I}_0$, from the arbitrarily positioned origin.  
In light of the prior choice
$\mathbf{I_{0}}=\mathbf{0}$, the probability adopts the canonical form $P \equiv P(\mathbf{I},i,N\, |\, i_0)$. 

During each time step of duration $\tau$, a particle located at a node adjacent to node $i$,
say, node $i'$, may enter node $i$ 
via a jump through an edge in $\Omega^{+}(i)$.  If the particle is located at node $i$, it may either 
remain there, say, due to a rejected jump caused by the presence of an obstacle, or exit through
one of the edges contained in $\Omega^{-}(i)$.  
For each edge $j$, denote the transition probability $w(j)$ 
that a particle located at the initial vertex of edge $j$ will chose to move through that edge
during time $\tau$, and for each node $i$, let $w(i)$ denote the probability that the particle will
choose to remain at node $i$ during time $\tau$.  Since a particle located at node $i$ must either remain there or exit
therefrom at 
each time step, 
there exists the conservation relationship \footnote{
In the context of the original discrete Taylor-Aris dispersion theory \cite{Dorfman:02}, it is possible to 
account for the rejected jumps by adding loops to the graph which return the particle to its original
position.  This alters the form of the master equation [cf. eq.  (\ref{Peqn2})], requiring an additional
sum over the outlet edges in $\Omega^{-}(i)$.  Since the underlying logic of the lattice random walk is
that the particle must make a jump at each time step (albeit with the possibility that this
jump will be rejected), the resulting calculation scheme is greatly simplified by requiring
a unitary probability [cf. eq.  (\ref{wconserve})]
of jumping and accounting for rejected jumps by the factor $w(i)$.
}:  
\begin{equation}
w(i) + \sum_{j \in \Omega^{-}(i)} w(j) = 1 \quad (\forall \ i). \label{wconserve}
\end{equation} 

The evolution of $P$ is governed by the following 
master equation quantifying the change in probability at 
each node on the global graph,
$\Gamma_g$, between time step $N$ and $N+1$:
\begin{eqnarray}
P(\mathbf{I},i,N+1)-P(\mathbf{I},i,N) = &&\sum_{j \in \Omega^{+}(i)} w(j)P(\mathbf{I'},i',N) \nonumber \\
                                        &&- [1-w(i)]P(\mathbf{I},i,N) \nonumber \\
						&& + \delta(\mathbf{I,0})\delta(i,i_0) \times \nonumber \\
						&&	\times \delta(N,0),
						\label{Peqn1}
\end{eqnarray}  
where the dependence upon the initial condition $i_0$ has been suppressed, and the $\delta(i,j)$ are Kronecker 
delta functions; $\delta(i,j) = 1$ if $i=j$, $\delta(i,j) = 0$ otherwise.

A rate equation may be obtained by dividing eq.  (\ref{Peqn1}) by the jump time, $\tau$.  
Interest here is focused upon the 
asymptotic, long-time behavior of $P$ after many jumps ($N \gg 1$), whence, to a good approximation \footnote{
It is expected that moving from a finite-difference approximation to a continuous derivative
reflects the physical process being modeled, since the lattice model merely represents a discrete 
approximation to a continuous process.},
\begin{equation}
\frac{P(\mathbf{I},i,N+1)-P(\mathbf{I},i,N)}{\tau} \approx \frac{dP(\mathbf{I},i,N)}{dt}.
\end{equation}
The time, $t \equiv N\tau$, is viewed as an essentially continuous variable for the case $N \gg 1$, which 
is equivalent to the long-time limit in conventional Taylor-Aris dispersion theory \cite{Dorfman:02}.
Use of the latter in eq.  (\ref{Peqn1}) furnishes the following differential algebraic equation
\begin{eqnarray}
\frac{dP(\mathbf{I},i,N)}{dt} = &&\sum_{j \in \Omega^{+}(i)} \left[\frac{w(j)}{\tau}\right]
					P(\mathbf{I'},i',N) \nonumber \\
                                        &&- \left[\frac{1-w(i)}{\tau}\right]P(\mathbf{I},i,N) \nonumber \\
						&&	+\tau^{-1} \delta(\mathbf{I})\delta(i,i_0)\delta(N). \label{Peqn2}
\end{eqnarray}  

\section{Moment Scheme}\label{Theory}
The master equation (\ref{Peqn2}) possesses a mathematical structure similar to the governing equation
appearing in Ref.  \onlinecite{Dorfman:02}, albeit with a significantly different physical interpretation.
Consequently, the moment-matching scheme employed previously may be invoked here to compute the mean velocity vector, 
$\bar{\mathbf{U}}^{\ast}$, and dispersivity dyadic, $\bar{\mathbf{D}}^{\ast}$, for 
the lattice transport problem defined by eq.  (\ref{Peqn2}).  The scheme is outlined in what follows, and the 
reader is referred to Ref.  
\onlinecite{Dorfman:02} for further details.

Define the local moment of the probability as the $m$-adic,
\begin{equation}
\mathbf{P}_m(i,N \, | \, i_0) \stackrel{\text{def.}}{=} \sum_{\mathbf{I}} 
									(\mathbf{R_I})^m P(\mathbf{I},i,N \, | \, i_0),
									\label{localdef}
\end{equation}
where $(\mathbf{R_I})^m \equiv \mathbf{R_I}\mathbf{R_I}\cdots\mathbf{R_I} $ ($m$ times) and the sum
over $\mathbf{I}$ represent the sum over all cells. (The spatial 
location of cell $\mathbf{I_{0}}=\mathbf{0}$ is $\mathbf{R}_{\mathbf{I}_0}=\mathbf{0}$.) The total moments
of the probability are the $m$-adics,
\begin{equation}
\mathbf{M}_m(N \, | \, i_0)\stackrel{\text{def.}}{=} \sum_{i \in V\Gamma_l} \mathbf{P}_m(i,N \, | \, i_0),
									\label{globaldef}
\end{equation}
where $V\Gamma_l$ are the vertices of the local graph. The zeroth moment is trivial,
\begin{equation}
M_0 = 1,
\end{equation}
which reflects the conservation of the probability on the lattice.  The next two higher-order total moments
are expected to 
grow linearly with time, and possess the respective forms:
\begin{eqnarray}
\frac{d\mathbf{M}_1}{dt} \approx \bar{\mathbf{U}}^{\ast}, \label{Ustardef} \\
\frac{d}{dt}\left(\mathbf{M}_2 - \mathbf{M}_1\mathbf{M}_1\right) \approx 2 \bar{\mathbf{D}}^{\ast}. \label{Dstardef}
\end{eqnarray} 

We briefly outline here the moment-matching scheme \cite{Dorfman:02} for computing 
$\bar{\mathbf{U}}^{\ast}$ and $\bar{\mathbf{D}}^{\ast}$:  (i) form the equations governing the local moments
from eqs.  (\ref{Peqn2}) and (\ref{localdef}); (ii) compute the asymptotic solutions to the first two local moments; 
(iii) form asymptotic total moments from the latter result and eq.  (\ref{globaldef}); and 
(iv) match the asymptotic moments of the exact solution to eqs.  (\ref{Ustardef})-(\ref{Dstardef}).  Only the
results of the moment scheme will be presented, since the mathematical manipulations required to arrive
at these results are lengthy and follow a procedure identical to that appearing elsewhere \cite{Dorfman:02}.

\subsection{Mean Velocity Vector}
The mean velocity vector is computed by forming the sum
\begin{equation}
\bar{\mathbf{U}}^{\ast} = \sum_{j \in E\Gamma_l, \ j \in \Omega^{+}}
				  \left[\frac{w(j)}{\tau}\right]\mathbf{R}(j)
				  P_0^{\infty}(i'), \label{Ustar}
\end{equation}
where $E\Gamma_l$ are the edges of the local graph, and the index 
$i'$ refers to the initial vertex of edge $j$.
The vertex field $P_0^{\infty}(i)$, corresponding to the asymptotic, steady-state probability of locating the 
particle at node $i$ in the unit cell,
is the solution of the linear set of equations,
\begin{equation}
\sum_{j \in \Omega^{+}(i)} w(j) P_0^{\infty}(i') 
- [1-w(i)] P_0^{\infty}(i) = 0. \label{P0}
\end{equation}
The latter are not linearly independent, and must be supplemented by the normalization condition
\begin{equation}
\sum_{i \in V\Gamma_l} P_0^{\infty}(i) = 1. \label{Pnormal}
\end{equation}

The present method of computing $\bar{\mathbf{U}}^{\ast}$ differs somewhat from that employed by 
Slater and coworkers (see, for example, Ref.  \onlinecite{Slater1}).  In the latter, the mean velocity is 
computed by first solving eqs.  (\ref{P0})-(\ref{Pnormal}) for 
$P_0^{\infty}(i)$.  
The velocity vector for each 
lattice site, say, $\mathbf{v}(i)$, is then defined as the difference in transition probability between
taking a forward step 
and a backwards step at site $i$, each step being weighted by its respective displacement (i.e.
zero displacement for rejected jumps).  The mean
velocity is then computed by the sum 
\begin{equation}
\bar{\mathbf{U}}^{\ast} = \sum_{i \in V\Gamma_l} \mathbf{v}(i)P_0^{\infty}(i).\label{SlaterU}
\end{equation}
Equations (\ref{Ustar}) and (\ref{SlaterU}) furnish identical values for $\bar{\mathbf{U}}^{\ast}$, since 
the difference between these two methods is simply a manifestation of Gauss' divergence theorem in a discrete
sense; eq.  (\ref{Ustar}) is the ``surface'' term and eq.  (\ref{SlaterU}) is the ``volume'' term.

\subsection{Dispersivity Dyadic}
The dispersivity dyadic is computed by the edge-sum
 \begin{equation}
\bar{\mathbf{D}}^{\ast} = \frac{1}{2}\sum_{j \in E\Gamma_l, \ j \in \Omega^{+}} \left[\frac{w(j)}{\tau}\right]
				  P_0^{\infty}(i') \tilde{\mathbf{b}}(j)\tilde{\mathbf{b}}(j), \label{Dstar}
\end{equation}
wherein 
\begin{equation}
\tilde{\mathbf{b}}(j) \stackrel{\text{def.}}{=} 
		\mathbf{R}(j) - \mathbf{B}(i) + \mathbf{B}(i'), \ j=\{i',i\} \label{btilde}
\end{equation}
for an edge directed from $i'$ to $i$.  The node-based vectors $\mathbf{B}(i)$ are 
the solutions of 
\begin{eqnarray}
  \sum_{j \in \Omega^{+}(i)} \left[\frac{w(j)}{\tau}\right] P_0^{\infty}(i') 
[\mathbf{B}(i') + \mathbf{R}(j) ] \nonumber \\
-\left[\frac{1-w(i)}{\tau}\right] P_0^{\infty}(i) \mathbf{B}(i) \nonumber \\
 =  P_0^{\infty}(i) \bar{\mathbf{U}}^{\ast}. \label{Beqn}
\end{eqnarray}
Equations (\ref{Beqn}) only define the $\mathbf{B}$ vectors to within an arbitrary, additive 
constant vector.  Consequently, this 
degree of freedom may be employed to set the $\mathbf{B}$ vector at one of the nodes, say $i^{\ast}$, equal to 
zero, $\mathbf{B}(i^{\ast})=\mathbf{0}$.

\section{Obstacle-Free Case}

Up to this point, no restrictions were placed upon the transition probabilities,
$w(j)$, the lattice spacing, $l$, or the jump time, $\tau$, aside from the conservation
statement (\ref{wconserve}), which ensures that the particle makes a decision (either
to jump to a new site or remain at the present site) during each time step.  However,
by considering the trivial case of transport on an obstacle-free lattice, 
additional relationships may be derived between the latter parameters, at least for the 
case of square lattices.

\begin{figure}[t]
\includegraphics[clip,width=2.5in]{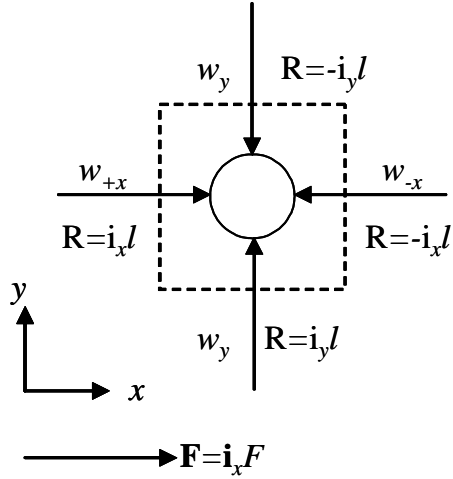}
\caption{
Basic graph (unit cell) for a lattice in the absence of obstacles.  The transition 
probabilities, $w(j)$, and macroscopic jump vectors, $\mathbf{R}(j)$
are indicated in the figure.  There are no rejected jumps, whereupon
$w(i)=0$. A force of magnitude $F$ is applied in the positive $x$-direction.
\label{Fig3a}
}
\end{figure}

Attention is restricted here
to a two dimensional lattice, whose basic graph is depicted in Fig. \ref{Fig3a}, since
generalizations to other dimensions are trivial.  
In the present circumstances, it only proves necessary to consider the case where the force
is directed in a single direction on the lattice, say, the $x$-direction,
since the spatial orientation of the lattice is aribtrary.  
The unit cell consists of a single
node with four entering edges, each of 
the latter possessing the transition probability
$w_{+x}$, $w_{-x}$ or $w_y$, depending on the edge orientation.

The governing equations (\ref{P0})-(\ref{Pnormal})
and (\ref{Beqn}) are satisfied here by the trivial solutions
$P_0^{\infty} = 1$ and $\mathbf{B} = \mathbf{0}$ [whereupon $\tilde{\mathbf{b}}(j)=\mathbf{R}(j)$].  From
eq. (\ref{Ustar}), the mean velocity adopts the form
\begin{equation}
\bar{\mathbf{U}}^{\ast} = \mathbf{i}_x \left(w_{+x}-w_{-x}\right) \frac{l}{\tau},
\end{equation}
and, from eq. (\ref{Dstar}), the dispersivity dyadic is
\begin{equation}
\bar{\mathbf{D}}^{\ast} =  \mathbf{i}_x \mathbf{i}_x \left[ \left(w_{+x}+w_{-x}\right)  \frac{l^2}{2 \tau} \right]
				 + \mathbf{i}_y \mathbf{i}_y \left( w_{y} \frac{l^2}{\tau} \right).
\end{equation}
For this trivial problem, though, the proper results are known {\em a priori}, namely $\bar{\mathbf{U}}^{\ast} = M \mathbf{F}$,
where $M$ is the free solution mobility and $\mathbf{F} = \mathbf{i}_x F$ is the applied force, and 
$\bar{\mathbf{D}} = \mathbf{I}D$, where $D$ is the molecular diffusivity and $\mathbf{I}$ is the idemfactor.
As a consequence, the usual velocity restriction \cite{Slater:86} is recovered,
\begin{equation}
\left(w_{+x}-w_{-x}\right) \frac{l}{\tau} = MF, \label{Urestrict}
\end{equation}
along with two restrictions governing the diffusive behavior,
\begin{equation}
\left(w_{+x}+w_{-x}\right) \frac{l^2}{2 \tau} = 
\frac{w_{y} l^2}{\tau} = D. \label{Drestrict}
\end{equation}

It is our contention that the latter restrictions (\ref{Urestrict})-(\ref{Drestrict}) must
be satisfied by any Monte Carlo algorith of this type, in order for the moments of the master 
equation to properly reflect the microscale physics.
For example, Gauthier and Slater \cite{Slater9} noted that their high-field
Monte Carlo algorithm, which furnishes the correct velocity (\ref{Urestrict}), predicts
that the obstacle-free diffusivity depends upon the 
field strength.  Indeed, it is readily 
confirmed that their choices of transition probabilities (and $\tau$) fail to satisfy eq.
(\ref{Drestrict}).  Consequently, the criteria of eq. (\ref{Drestrict}) serves as 
a litmus test for newly proposed algorithms.

If the following normalization condition is enforced,
\begin{equation}
w_{+x} + w_{-x} + 2w_{y} = 1, \label{wnormal}
\end{equation}
then eqs. (\ref{Urestrict})-(\ref{wnormal}) constitute 4 equations
for 5 unknowns.  If the lattice spacing, $l$, is left as an ajustable parameter,
then one recovers the usual ``small-bias'' algorithm (see, for example, Ref. \onlinecite{Slater1})
\begin{equation}
w_{\pm x} = \frac{1 \pm \epsilon}{4}, \quad 
w_{y} = \frac{1}{4}, \quad 
\tau = \frac{l^2}{4D}, \label{parameters}
\end{equation}
where $\epsilon$ is the reduced field,
\begin{equation}
\epsilon \overset{\text{def.}}{=} \frac{Fl}{2kT}. \label{epsilon}
\end{equation}
Remarkably, the latter results were derived solely from considering the obstacle-free velocity
and diffusion coefficient, along with the normalization condition (\ref{wnormal}), without
ever explicitly stating that $\epsilon \ll 1$.  However, the transition probabilities
are only sensible in the latter limit, or at least for the inequality $\epsilon < 1$, since
the physical interpretation of a negative transition probability is suspect.

\section{Transport on an Isotropic Lattice in a Finite External Field}

\begin{figure}[t]
\includegraphics[clip,width=2.5in]{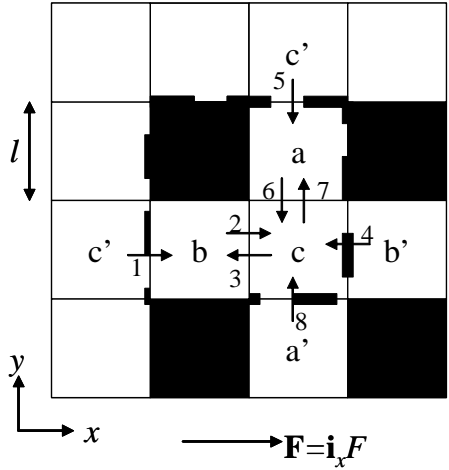}
\caption{Isotropic array of square obstacles of characteristic length $l$ separated by a distance $l$.  
The repetitive unit cell is indicated by the dashed lines. An external force 
of strength $F$ is applied in the $x$-direction to animate the particle.  The labeled
lattice sites and indicated arrows constitute the basic graph, $\Gamma_{b}$.
\label{Fig3}
}
\end{figure}

In the following illustrative example, we analyze transport occurring 
in a simple isotropic system, depicted in Fig. \ref{Fig3}, under the influence of a finite field.  
The lattice, consisting of square obstacles of size $l$ separated by a distance $l$, constitutes
the simplest configuration for demonstrating 
the present scheme.  

In the parlance of the present theory,  and with
the edge numbering depicted in Fig. \ref{Fig3}, the transition probabilities are
$w(j) = w_{+x}$ for $(j=1,2)$, $w(j) = w_{-x}$ for 
$(j=3,4)$, and $w(j) =w_{y}$ otherwise.  For now,
the specific forms of the transition probabilities, $w(j)$, and the jump time, $\tau$, will be
left unspecified, although subject to the generic physical restrictions (\ref{Urestrict})-(\ref{Drestrict}).
  As a consequence of rejected jumps, the probabilities of 
remaining at the different lattice sites are
$w(a) = w_{+x}+w_{-x}$, $w(b) = 2w_{y}$ and $w(c) = 0$.  
The problem specification is completed by the macroscopic jump vectors,
\begin{equation}
\mathbf{R}(j) = 2l \left\{
			\begin{array}{rl}
			 \mathbf{i}_x, 	&	j = 1, \\
			- \mathbf{i}_x,	&	j = 4, \\
			- \mathbf{i}_y,	&	j = 5, \\
			 \mathbf{i}_y,	&	j = 8, \\
			\mathbf{0},		&	\text{otherwise}.
			\end{array}
		    \right. 
\end{equation}

It is readily apparent that the solution $P_0^{\infty} = 1/3$ satisfies eqs.  (\ref{P0})-(\ref{Pnormal}), 
a consequence of the isotropy of the lattice.  Substituting the relevant parameters into
eq.  (\ref{Ustar}) furnishes the mean velocity,
\begin{equation}
\bar{\mathbf{U}}^{\ast} = \frac{2}{3} \left(w_{+x} - w_{-x} \right) \frac{l}{\tau}\mathbf{i}_{x} , 
\end{equation}
which, with use of eq. (\ref{Urestrict}), adopts the form
\begin{equation}
\bar{\mathbf{U}}^{\ast} = \frac{2}{3}M \mathbf{F}. \label{Ustar1}
\end{equation}
The latter result is invariant to the (arbitrary) choice 
of axis labels,
whereupon it is recognized that isotropy of the lattice geometry renders the mean mobility tensor isotropic,
\begin{equation}
\bar{\mathbf{M}}^{\ast}	 = \mathbf{I} \bar{M}^{\ast}, 
\quad \bar{M}^{\ast} = \frac{2}{3}M. \label{Ustar2}
\end{equation}
The effective mobility, $\bar{M}^{\ast}$, is less than the free solution value, $M$, owing to  
the inability to make forward jumps on lattice site $a$.
The latter observation, as well as the equivalence of eqs.  (\ref{Ustar}) and (\ref{SlaterU}),
is rendered transparent by this considering the local velocity \cite{Slater2} at each node,
\begin{equation}
\mathbf{v}(i) = \left\{ \begin{array}{ll}
				M\mathbf{F}, 	&	i = b,c, \\
				\mathbf{0},						&	i = a.
			\end{array}\right.
\end{equation}
Use of the latter and eq.  (\ref{SlaterU}) furnishes the result (\ref{Ustar1}).

Since nodes $a$ and $b$ communicate via node $c$, the remaining calculations are greatly simplified
by choosing
$\mathbf{B}(c)=\mathbf{0}$, whereupon eq.  (\ref{Beqn}) 
directly furnishes the values of the remaining $\mathbf{B}$ vectors,
\begin{eqnarray}
\frac{\mathbf{B}(a)}{l} & = & \frac{2}{3} \left( \frac{w_{-x}-w_{+x}}{2 w_{y}}\right) \mathbf{i}_{x} - \mathbf{i}_{y}, \label{Ba} \\
\frac{\mathbf{B}(b)}{l} & = & \frac{2}{3}\left(\frac{2w_{+x} + w_{-x}}{w_{x} + w_{-x}}\right)\mathbf{i}_{x}, \label{Bb}
\end{eqnarray}
which have been simplified with use of eq.  (\ref{wnormal}).

Compute $\tilde{\mathbf{b}}(j)$ from eq.  (\ref{btilde}), with use of eqs.  (\ref{R1}) and (\ref{Bsoln1}), 
and use the result in eq.  (\ref{Ddiff}) to ultimately arrive at the dispersivity dyadic
\begin{equation}
\frac{\bar{\mathbf{D}}^{\ast}}{D} = \bar{D}^{\ast}_{xx} \mathbf{i}_x \mathbf{i}_x 
						+ \bar{D}^{\ast}_{yy} \mathbf{i}_y \mathbf{i}_y, 
\end{equation}
where the components of the dispersivity,
\begin{eqnarray}
\frac{\bar{D}^{\ast}_{xx}}{D} & = & \frac{4}{27} \left[\begin{array}{l} 
									 \frac{\left(2w_{+x}+w_{-x}\right)^2 + 
									\left(2w_{-x}+w_{+x}\right)^2}{\left(w_{+x}
									+w_{-x}\right)^2} + \\
								 + \frac{\left(w_{+x}-w_{-x}\right)^2}{2w_{y}^{2}} \\ 
									\end{array} \right], \label{Dxx} \\
\frac{\bar{D}^{\ast}_{yy}}{D} & = & \frac{2}{3}, \label{Dyy}
\end{eqnarray}
have been rendered in terms of the molecular diffusivity, $D$, with use of eq. (\ref{Drestrict}).

Clearly, the lateral dispersivity, $\bar{D}^{\ast}_{yy}$, obeys the Nernst-Einstein relationship
(\ref{Nernst}) for any choice of the transition parameters, as must be the case since there exists no bias in the
lateral direction.  In contrast, the axial dispersivity, $\bar{D}^{\ast}_{xx}$, only obeys the 
Nernst-Einstein relationship in the limit where $w = 1/4 \ (\forall \ j)$, i.e. for pure 
molecular diffusion.  Importantly, the axial dispersion is also invariant to the arbitrary orientation of the
$x$-axis.
If one adopts the small-bias parameters (\ref{parameters}), the axial dispersion
reduces to the form
\begin{equation}
\frac{\bar{D}^{\ast}_{xx}}{D} = \frac{2}{3} + \frac{10}{27}\epsilon^2.
\end{equation}
For finite field strengths, there exists a so-called ``Taylor'' contribution (or convective dispersivity)
in the direction of the field, the latter depending quadratically upon $\epsilon$, or, equivalently,
the square of the Peclet number.  The latter result agrees with conventional continuum analyses of
the convective dispersivity.

\section{Unbiased Random Walk}\label{Unbiased}
\subsection{Simplification of the General Theory}
Significant simplifications are realized when $w(j) = w = \text{const.} \ (\forall \ j)$, which
is equivalent in its consequences to an unbiased random walk (or 
molecular diffusion) on the lattice, albeit with the possibility of rejected steps due to the presence
of the obstacles.
For a unit cell with $\eta$ available sites, the asymptotic probability adopts the form
\begin{equation}
P_0^{\infty}(i) = \eta^{-1} \ (\forall \ i), \label{Pdiff}
\end{equation}
reflecting the fact that it is equally likely for the particle to be located at any of the available sites \footnote{
When using the Nernst-Einstein relationship (\ref{Nernst}) to compute the effective diffusivity in the zero-field limit,
the presence of an infinitesimally small force induces a bias in $P_0^{\infty}$.  (As an illustrative example, 
consider the effect of curved field lines \cite{Slater8}.)  The latter bias
appears here only in the $\mathbf{B}$-field, since the force is identically zero.
}.  
With use of eq.  (\ref{wconserve}), it is easily shown that eq.  (\ref{Pdiff}) satisfies eqs.  (\ref{P0})-(\ref{Pnormal}).
Substitution of eq.  (\ref{Pdiff}) into (\ref{Ustar}) reveals that
\begin{equation}
\bar{\mathbf{U}}^{\ast} = \left[\frac{w}{\eta \tau}\right] 
				\sum_{j \in E\Gamma_l } \mathbf{R}(j). \label{Udiff}
\end{equation}
Since $E\Gamma_l$ only contains edges entering the unit cell, the lattice structure guarantees that 
each edge possessing the 
macroscopic jump vector $\mathbf{R}(j)$ can be paired with another edge whose macroscopic
jump vector is $-\mathbf{R}(j)$ (see, for example, Fig. \ref{Fig2}).  Consequently, the summation 
appearing in eq.  (\ref{Udiff}) is identically zero, whereupon 
\begin{equation}
\bar{\mathbf{U}}^{\ast}=\mathbf{0},
\end{equation}
as must be the case for a pure diffusion problem.

The $\mathbf{B}$-equation (\ref{Beqn}) undergoes a similar simplification in the unbiased limit,
\begin{equation}
 \sum_{j \in \Omega^{+}(i)} [\mathbf{B}(i') + \mathbf{R}(j) ] 
 - \left[\frac{1-w(i)}{w}\right]
\mathbf{B}(i) = \mathbf{0}. \label{Bdiff}
\end{equation}
The $\tilde{\mathbf{b}}(j)$ are still computed by eq.  (\ref{btilde}), whereupon $\bar{\mathbf{D}}^{\ast}$ 
adopts the form 
\begin{equation}
\bar{\mathbf{D}}^{\ast} = \left[\frac{w}{2\eta\tau}\right] \sum_{j \in E\Gamma_l }
				  \tilde{\mathbf{b}}(j)\tilde{\mathbf{b}}(j), \label{Ddiff}
\end{equation}

For certain geometries, the present scheme for computing the effective diffusivity is 
computationally more efficient than 
the alternative procedure \cite{Mercier:99} of evaluating $\bar{\mathbf{M}}^{\ast}$ 
in the low-force limit and then using eq.  (\ref{Nernst})
to compute $\bar{\mathbf{D}}^{\ast}$.  First, the 
coefficient matrix appearing in eq.  (\ref{Bdiff}) is invariant to 
spatial direction.
Consequently, it is only necessary to specify the new solution vector in order to 
calculate a second (or third)
component of $\mathbf{B}$.  In contrast, the method of Ref.  \onlinecite{Mercier:99} requires computing 
a separate set of probabilities corresponding to the external force being 
oriented in each direction, each calculation (generally) involving a new 
coefficient matrix. While both methods require the same number of matrix inversions, the $\mathbf{B}$
matrices are one dimension smaller.  However, we expect this slight edge to be offset by the more
involved calculations required to compute $\bar{\mathbf{D}}^{\ast}$ by eq.  (\ref{Ddiff}).  Importantly, though, 
the present method corresponds {\em exactly} to the case of no
applied force, which eliminates the need to obtain the 
perturbation to the uniform probability distribution in 
the low-force limit, although this can be done
fairly efficiently \cite{Slater4}.  The Nernst-Einstein equation scheme is preferable, however, when significant
geometrical simplifications can be effected by eliminating the macroscopic jump vectors and 
making use of ``identical'' nodes.
That being said, both the present technique and the Nernst-Einstein
limit technique are markedly less complicated than first-passage time analyses, and either method
serves as an efficient analytical alternative to conventional simulation techniques, as well as 
a check on complex numerical codes.

\subsection{Pure Diffusion in an Asymmetric Array}\label{Mercier}
As a final example, we
compute the effective diffusivity
of a particle moving on the lattice shown in Fig. \ref{Fig1}.  The effective diffusivity for this
particular geometry was previously computed by 
Mercier {\em et al.} \cite{Mercier:99} using the techniques mentioned \S\ref{Intro}, and we 
demonstrate here that the present theory reproduces their results.

There are
seven available locations in the unit cell, $\eta = 7$, and the jump probability is $w = 1/4$. 
The jump time, $\tau$, is
given by the Brownian time scale of eq. (\ref{parameters}).
The probability of remaining on node $i$ is equal to 
the number of adjacent unavailable sites divided by four,
\begin{equation}
w(i) = \left\{
			\begin{array}{ll}
			0,	&	i = b,c,d,f, \\
			1/4,  	&	i = e,g, \\
			1/2,	&	i = a.
			\end{array}
	\right.
\end{equation}
With the edge
numbers depicted in Fig. \ref{Fig2}, the macroscopic jump vectors are
\begin{equation}
\mathbf{R}(j) = 2l \left\{
			\begin{array}{rl}
			2 \mathbf{i}_x, 	&	j = 1, \\
			-2 \mathbf{i}_x,	&	j = 2, \\

			- \mathbf{i}_y,	&	j = 3-5, \\
			 \mathbf{i}_y,	&	j = 6-8, \\
			\mathbf{0},		&	\text{otherwise}.
			\end{array}
	\right. \label{R1}
\end{equation}

Substituting the relevant parameters into eq.  (\ref{Bdiff}) and  choosing $i^{\ast}=d$ 
(i.e. $\mathbf{B}(d) = \mathbf{0}$), 
\begin{eqnarray}
&&
\left[\begin{array}{cccccc}
	-2  & 1 & 0 & 0 & 0 & 0 \\
	1 & -4 & 1 & 2 & 0 & 0 \\
	0 & 1 & -4 & 0 & 2 & 0 \\
	0 & 2 & 0 & -3 & 1 & 0 \\
	0 & 0 & 2 & 1 & -4 & 1  \\
	 0 & 0 & 0 & 0 & 1 & -3 
	\end{array}
\right]\cdot
\left[\begin{array}{c}

	\mathbf{B}(a) \\
	\mathbf{B}(b) \\
	\mathbf{B}(c) \\
	\mathbf{B}(e) \\
	\mathbf{B}(f) \\
	\mathbf{B}(g) 
	\end{array}
\right] \nonumber \\
&& =
-4l \left[\begin{array}{c}
	1 \\
	0 \\
	0 \\
	0 \\
	0 \\
	0 
	\end{array}
\right] \mathbf{i}_x
+
2l \left[\begin{array}{c}
	0 \\
	1 \\
	1 \\
	-1 \\
	-1 \\
	-1 
	\end{array}
\right] \mathbf{i}_y.
\end{eqnarray}

The solution of this system of equations is
\begin{equation}
\left[\begin{array}{c}
	\mathbf{B}(a) \\
	\mathbf{B}(b) \\
	\mathbf{B}(c) \\
	\mathbf{B}(e) \\
	\mathbf{B}(f) \\

	\mathbf{B}(g) 
	\end{array}
\right] \\
=
\frac{l}{4} \left[\begin{array}{c}
	11 \\
	6 \\
	3 \\
	5 \\

	3 \\
	1 
	\end{array}
\right] \mathbf{i}_x
+
l \left[\begin{array}{c}
	0 \\
	0 \\
	0 \\
	1 \\
	1 \\
	1 
	\end{array}
\right] \mathbf{i}_y. \label{Bsoln1}
\end{equation}
Compute $\tilde{\mathbf{b}}(j)$ from eq.  (\ref{btilde}), with use of eqs.  (\ref{R1}) and (\ref{Bsoln1}), 
and use the result in eq.  (\ref{Ddiff}) to furnish the dispersivity dyadic
\begin{equation}
\bar{\mathbf{D}}^{\ast} = \left(\frac{5}{28} \mathbf{i}_x\mathbf{i}_x +
				  \frac{3}{14} \mathbf{i}_y\mathbf{i}_y\right) \frac{l^2}{\tau},
\end{equation}
or, in terms of $D$,
\begin{equation}
\bar{\mathbf{D}}^{\ast} = \left(\frac{5}{7} \mathbf{i}_x\mathbf{i}_x +
				  \frac{6}{7} \mathbf{i}_y\mathbf{i}_y\right) D,
\end{equation}

The latter result agrees exactly with Ref.  \onlinecite{Mercier:99}.  
Although not shown here, the present calculation scheme also reproduces the 
low-field mobilities obtained previously \cite{Mercier:99} during the course of the Nernst-Einstein calculation,
although knowledge of the latter mobilities proves unnecessary to compute
$\bar{\mathbf{D}}^{\ast}$.

Having furnished a third independent verification of the value of $\bar{\mathbf{D}}^{\ast}$ for this array,
we turn our attention to the volume-averaging theory \cite{Keller:02} and investigate its comparable result.
The pure diffusion problem includes no
bias in the transition probabilities and results in zero mean velocity, whereupon
their generic result for the $yy$ component of the dispersivity adopts the form
\begin{equation}
\bar{D}^{\ast}_{yy} = \Gamma \left( 1 - \sum_{W} \hat{n}_y g^{y} \right) - v_y \sum g^y, \label{Dyywrong}
\end{equation}
where $\Gamma$ is their dimensionless diffusivity, $\sum_{W}$ is the summation over all sites adjacent
to the obstacle, $\hat{n}_y$ are outward pointing normal vectors from the obstacles, 
$\sum$ is the sum over all available sites, $v_y$ is the
microscopic drift velocity, and $g^y$ is the volume-averaging quantity which plays a similar role to 
our $\mathbf{B}$ field.
Refering back to Fig. \ref{Fig1}, the quantity $v_y = 0$, since the transport process only includes
diffusion in the $y$-direction.  Consequently, the second term in eq. (\ref{Dyywrong}) makes no contribution
to the effective diffusivity. 
Moreover, only site $a$ has a non-zero value of $\hat{n}_y$. Indeed, the latter site contains both 
a positive and negative unit vector, corresponding to the lower and upper obstacle sites.  As such, the 
summation $\sum_{W}$ makes no contribution to $\bar{D}^{\ast}_{yy}$, and one recovers the physically incorrect
result $\bar{D}^{\ast}_{yy} = \Gamma = D$.

\section{Concluding Remarks}
The present contribution has developed a generic scheme for computing the mean velocity vector,
$\bar{\mathbf{U}}^{\ast}$, 
and dispersivity dyadic, $\bar{\mathbf{D}}^{\ast}$,
from lattice models of transport in spatially periodic arrays of obstacles, the latter representing simple
models of gel electrophoresis.  To the best of our knowledge, this constitutes the only
analytical method available at the present time for correctly computing the dispersivity dyadic from 
lattice Monte Carlo models for  finite fields.  
The power and simplicity of the present calculation scheme renders it 
appropriate for revisiting the Ogston sieving problems considered previously
\cite{Slater1,Slater2,Slater3,Slater4,Slater5,Slater6,Slater7,Slater8,Slater9}.  
in order to investigate the non-trivial solute dispersion arising during 
such transport processes.  

Furthermore, it may be possible to use the general framework developed here to analyze
DNA separations in microfluidic devices \cite{Chou:99,Cabodi:02,Doyle:02}, provided that rational
rules can be established for the transition probabilities prevailing therein.  
In the case of so-called 
rectified Brownian motion \cite{Chou:99,Cabodi:02,Duke:98} (or vector chromatography \cite{Dorfman:01}),
a directional separation is achieved by applying the force at an angle with 
respect to the symmetry axes of the array.  One should be able to analyze these devices in a manner similar
to Ref. \onlinecite{Keller:02}, although it is recommended that the present approach
be employed to compute the dispersivity.  It would be interesting to consider
a mesh which is sufficiently fine in order to account for 
local variations in field strength and direction arising from the fact that the electric 
field does not penetrate the obstacles.  The latter variations were
recently demonstrated to have a strong effect on the experimental realization of directional separation schemes
 \cite{Huang:02}.

In the case of magnetosensitive
arrays \cite{Doyle:02}, the separation is achieved 
by a size-specific ``hold-up'' of the different DNA strands on the posts of the array.  
Consequently, application of the present scheme would require augmenting the 
hard-core repulsion of the obstacles to also account for the ``attractive'' residence time of the molecules
when they become hooked.  This problem is analogous to the inclusion of attractive forces exerted by the fibers of the
gel \cite{Slater5}, although exactly defining what is meant by the ``attraction'' to the posts
is somewhat more ambiguous, since the attraction is not caused by 
a physical potential well.

\begin{acknowledgments}
The author would like to thank Gary Slater of University of Ottowa for 
an enlightening discussion on Monte Carlo simulations. 
The encouragement
of Jean-Louis Viovy of Institut Curie is greatly appreciated.
This work was supported by a Postdoctoral Fellowship from Institut Curie.
\end{acknowledgments}

\end{document}